\documentclass{article}

\usepackage{arxiv}

\usepackage[utf8]{inputenc} 
\usepackage[T1]{fontenc}    
\usepackage{hyperref}       
\usepackage{url}            
\usepackage{booktabs}       
\usepackage{amsfonts}       
\usepackage{nicefrac}       
\usepackage{microtype}      
\usepackage{lipsum}		
\usepackage{graphicx}
\usepackage[numbers]{natbib}
\usepackage{doi}

\usepackage{times}
\usepackage{graphicx}
\usepackage{amssymb}
\usepackage{mathtools}
\usepackage{nth}
\usepackage{multirow}
\usepackage{booktabs}
\usepackage{bm,amsmath}
\usepackage{url,hyperref}

\title{Identifying and examining machine learning biases on Adult dataset}

\author{ 
    Sahil Girhepuje \thanks{Author can be contacted at \texttt{www.linkedin.com/in/sahil-girhepuje/}} \\
    Department of Engineering Design \\
    Indian Institute of Technology Madras\\
    Chennai - 600036 \\
    \texttt{ed19b048@smail.iitm.ac.in}  \\
}

\date{}

\renewcommand{\headeright}{Identifying and examining machine learning biases on Adult dataset}

\renewcommand{\undertitle}{}
\renewcommand{\shorttitle}{}

\hypersetup{
pdftitle={Identifying and examining machine learning biases on Adult dataset},
pdfsubject={q-bio.NC, q-bio.QM},
pdfauthor={Sahil Girhepuje},
pdfkeywords={Machine Learning Bias, Fairness Assessment},
}

\begin{document}
\maketitle
.

\begin{abstract}
	This research delves into the reduction of machine learning model bias through Ensemble Learning. Our rigorous methodology comprehensively assesses bias across various categorical variables, ultimately revealing a pronounced gender attribute bias. The empirical evidence unveils a substantial gender-based wage prediction disparity: wages predicted for males, initially at \$902.91, significantly decrease to \$774.31 when the gender attribute is alternated to females. Notably, Kullback-Leibler divergence scores point to gender bias, with values exceeding 0.13, predominantly within tree-based models. Employing Ensemble Learning elucidates the quest for fairness and transparency. Intriguingly, our findings reveal that the stacked model aligns with individual models, confirming the resilience of model bias. This study underscores ethical considerations and advocates the implementation of hybrid models for a data-driven society marked by impartiality and inclusivity.
\end{abstract}



\section{Introduction}
The University of California (UCI) Adult dataset is iconic in the Machine Learning domain. It is extensively used to benchmark new algorithms. The dataset is considered a good starter example to implement novel techniques \cite{Chakrabarty2018} Intelligence (AI) and Computer Science that focuses on using data and algorithms to imitate how human beings learn. ML has been used in automatic translation, image recognition and spam classification \cite{Taniguchi2018} technological applications. It enables computers to perform these tasks with remarkable accuracy and great speed. AI often gives the impression that it is objective and fair. However, algorithms are made by humans and trained by data which may be biased. In this background, a question may arise - \emph{Are ML models free from biases?} There have been several examples of deployed AI algorithms that have made biased decisions - even when there was no intention for it \cite{Gade2019, Piech} . In this regard, can we identify techniques to reduce the bias? The literature suggests that ML Bias is a central topic of discussion for Artificial Intelligence. This topic attracts researchers' attention due to the biased nature of algorithms used in ML models. Research in this direction is essential for the future of Artificial Intelligence. 

ML models are built using training data collected from human beings' experiences about worldly affairs. Human beings possess a cognitive bias in their thinking and behaviour patterns \cite{Friedman2017} . Since ML models mainly replicate human beings- thinking and actions, these models also contain biases \cite{Alelyani2021} . Therefore, the results may be biased when a biased ML model is processed on a dataset. Hence, we can treat it as a "bias in and bias out" system.\\
Because of biases inherent in the training data, addressing discriminatory algorithmic judgements remains a challenge \cite{Starke2021} . Questions and concerns are raised about using biased algorithms in the decision-making process. Identifying and evaluating algorithm biases is essential for developing and designing better ML models. A few research studies have shown that ML models suffer from biased results by quantifying levels of biases. However, quantitative analyses on biased ML models are not examined and analysed in detail.

The present research study analyses the descriptive ML biases issues and finds out the effects of stacking of ML models while evaluating the ML biases. To carry out the research tasks, we use alternation functions and Potentially Biased Attributes (PBAs). We evaluate and elucidate the impact on prediction using KL divergence. 
This study will 
\begin{itemize}
    \item explore the Potentially Biased Attributes (PBAs) in the Census dataset
    \item examine the effects of applying alternation functions on the PBAs
    \item evaluate the levels of bias using KL divergence
    \item analyse the impact of using stacking of ML models on the biases
\end{itemize}


\section{Related Work}
Human beings' cognitive bias is an extensively-studied fact in psychology \cite{Haselton2015} . The cognitive bias in human beings' judgement has shown to promote fast learning \cite{Taniguchi2018}. For decision-making, an enormous amount of information in a human's brain is filtered, and only relevant information is used. 

The terms bias and unfairness are often interchangeably used in everyday use. Nevertheless, researchers believe bias does not imply unfairness. In the real world, unfairness can be introduced by altering a specific criteria to prefer a particular group over another. The criteria in ML are seen as algorithmic parameters. Thus, unfairness is inducted into the algorithm, while bias naturally exists in data. Similarly, \citet{Agarwal2018}  argue that the bias is introduced by humans responsible for labelling the training data. Data labelling refers to adding meaningful and informative labels to each sample in the raw data. Data labelling provides a context so that an ML model can learn from it. The authors also claim that humans are not the only sources of bias, but algorithms are as well. For instance, recommender systems that provide humans with their preferred content can be seen as additional sources of bias.  

Alelyani's \cite{Alelyani2021} literature review mentions that cognitive bias is not always bad. They propose that cognitive bias may advance the decision-making process. However, the important theme which emerges from studies is that regardless of the benefit of human cognitive bias, ML bias is undesirable. This is because it impacts the decision-making ability of the algorithm. \citet{McCradden2020} discuss how improvements in Patient safety can be brought about if the ML bias in healthcare is controlled. 

Several studies claim that ML bias comes from the training data. For instance, the  degree attribute for job applicants might be biased due to the indirect impact of the gender attribute. Only a few females major in technology, and even fewer have graduate degrees in technology-related fields, such as engineering. In this case, it is the gender attribute which introduces bias in the learning model, not the degree attribute \cite{Alelyani2021}. Although auditing of algorithms has emerged as a significant strategy to uncover systematic biases embedded in software platforms, researchers struggle to understand the real-world impact of these audits \cite{ActionableAuditing2019} 

\begin{figure*}
    \begin{center}
        \includegraphics[width=4in]{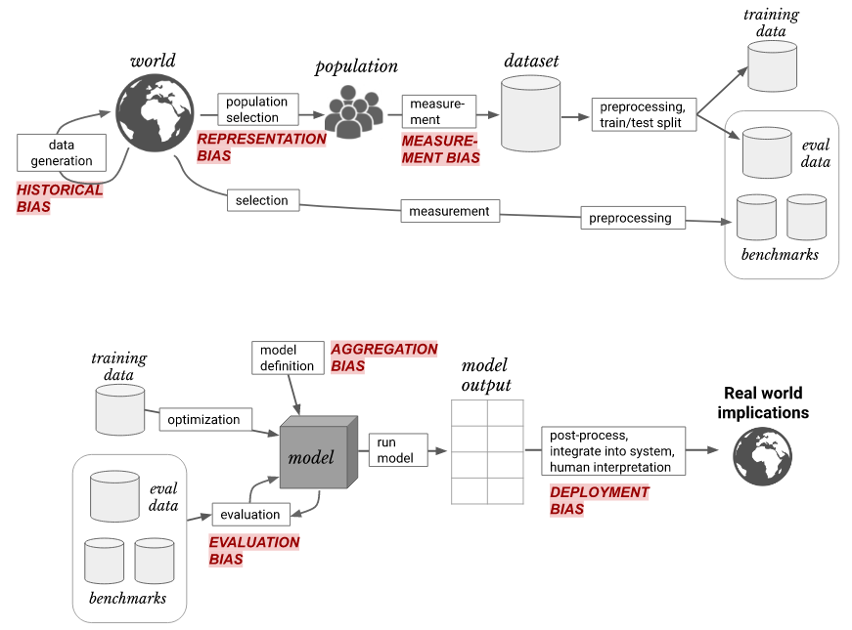}
    \end{center}
    \caption{Several systematic reviews of Machine Learning Bias have been undertaken. In "A Framework for Understanding Unintended Consequences of Machine Learning", MIT's authors \citet {Suresh2021} describe six types of bias in machine learning, summarised in their paper. They show how each step in the ML model-building process can contribute to bias. Aspects of data generation, like the selection of population and measurement of responses, can also add to the biases. The method of defining the ML model, its evaluation and deployment - any biases here lead to real-world implications}
    \label{fig: mit bias}
\end{figure*}

Much of the research has focused on identifying the types of bias. The work of \citet{Kochling} is another good summary. They perform an organised review of 36 journal articles from 2014 to 2020. They present some applications of algorithmic decision-making and evaluate the possible pitfalls. Twenty types of cognitive biases and possible debiasing techniques are covered by \citet{Kliegr2021}. The authors demonstrate that designers of ML algorithms and software can adapt their research. Biased results are called explainable if these attributes can justify the differences in treatment among distinct groups \cite{Mehrabi2021}. Few types of bias emerge from previous studies: information bias, algorithmic bias, representation bias, aggregation bias, and mutually exclusive bias. For our research, we limit our work to representation and aggregation bias. 

In a significant study, \citet{Maria2019} mention a gender imbalance in occupations . The researchers notice that models predicting occupation not only reflect the gender imbalance, but also amplify it. For example, in their training data about surgeons, 14.6\% of surgeons are women. However, when an ML model was trained to predict the gender of a surgeon, it was found that only 11.6\% of the true positives were women. The model thus amplified the bias existing in the training set. This representation bias is quite common, particularly for simple ML models such as Linear Regression. Aggregation bias occurs when models do not aggregate data to incorporate all relevant factors  \cite{Howard2020} . For instance, the treatment of diabetes is often based on studies involving small groups of heterogeneous people. Usually, the analysis of results is done in a way that does not consider different ethnicities or genders. However, it turns out that diabetes patients have various complications across ethnicities \cite{Spanakis2013} . The HbA1c levels, which are widely used to diagnose and monitor diabetes, differ in complex ways across ethnicities and genders \cite{Herman2012}. Since medical decisions are based on a model that does not include these crucial variables and interactions,  this can result in severe misdiagnosis or incorrect treatment. Citing similar arguments, the authors in "Towards a pragmatist dealing with algorithmic bias in medical machine learning" say - "we thus plead for a pragmatist dealing with algorithmic bias in healthcare environments" \cite{Starke2021}.

A detailed examination by \citet{Alelyani2021} showed that we could eliminate protected attributes from training data to wipe out bias. However, in the real world, we cannot simply assume the existence of bias without detecting it first. \citet{Solanes2021} draw our attention to similar circumstances in the medical domain. They show that authors of novel machine-learning MRI studies may remove the EoS attribute (EoS is a term in medical imaging) when training the ML models. But they may not control EoS when estimating their model's accuracy. This potentially leads to severely biased estimates. They discuss how further control of the EoS attribute is required for a non-inflated result. Another example of working with biased data is shown by \citet{Kumar2019}  . They use Karl Max's and Ricardo Solow's hypotheses to analyse India's Income by using limited data, hence intentionally introducing \emph{bias} in training. 

The notion of using multiple ML models to predict an outcome is outlined in "Real-time prediction of online shoppers' purchasing intention using multi-layer perceptron and LSTM recurrent neural networks"  \cite{Sakar2019}. To increase purchase conversion rates, the researchers use two modules. The first module consists of random forest (RF), support vector machines (SVMs), and multi-layer perceptron  (MLP)  classifiers. After suitable data-processing steps, the second module is used. This module consists of long short-term memory (LSTM) based on recurrent neural networks. The output from the second module shows the probability estimate of the visitor's intention to leave the site without finalising the transaction. 

While \citet{Sakar2019} focus on getting better predictions, \citet{Lundberg} are more concerned with making an ML model explainable. Many ML systems are considered \emph{black boxes} as it is hard to gain a comprehensive understanding of their inner working after they have been trained. In other words, understanding why a model makes a specific prediction can be as crucial as the prediction's accuracy in many applications. However, the highest accuracy for modern datasets is often achieved by complex models that even experts struggle to interpret, creating a tension between accuracy and interpretability. One of the most influential accounts of model readability comes from \citet{Lundberg}. SHAP (SHapleyAdditive exPlanations). SHAP assigns each feature an importance value for a particular prediction. The view of explainable models is supported by \citet{You2015} . Using ML algorithms, they use CHAID decision trees and Pareto values to identify important attribute values of groups made. 

\citet{Reigada2021} emphasises the importance of the dataset in use. He cautions against \emph{blindly following the dataset} by saying that a given dataset may not be complex enough to predict the required output value. For instance, there are various reasons why a customer would respond to marketing. The dataset used may only include a small fraction of all variables that should be considered. 

\citet{Nham2019} analyses the properties of the Census dataset (discussed below) and uses classification algorithms like logistic regression, Naive Bayes, and decision trees to make predictions on the income. More importantly, it raises the crucial question of how different attributes correlate with one another. For instance, \emph{of the individuals who make more than \$50,000 in a year, does their race affect their work hours?}. He stresses the role of Exploratory Data Analysis steps (EDA) when working with an ML model. We find an exciting finding in his work -if we predict high income for all-male bachelors holding a bachelor's degree, then this 'model' surpasses the accuracy score by other ML models. Such caveats can only be rejected if we know from the exploratory analysis that the data is biased toward white males with college degrees. 

Taken together, these studies support the notion that if we want to eliminate or mitigate model bias, we need to tune the model's parameters to obey our desires of expected outcomes \cite{Bellamy2018}. Tuning the parameters is equivalent to changing the hiring criteria to accredit one gender over another, which is not fair \cite{Alelyani2021, Rajkomar2018} . Research consistently points towards the fact that all datasets contain bias. We also saw how Machine Learning could amplify bias because human bias can lead to more significant amounts of ML bias. As practitioners of ML, we should step back for a moment and consider: How would we feel if we discovered that we had been part of a system that ended up hurting society? \cite{Kochling}
The problem is: How can we be sure about the presence of bias until we detect it and quantify it \cite{Alelyani2021, Barbu2019}.


\section{Methodology}
To attain the research project's objectives and to answer the research questions we have adapted, we use analytical and quantitative research methodology to carry out research. This is because the basis for this research has been confirmed in previous studies. We critically evaluate and analyse their findings. This involves statistical training to analyse the data. Further, we aim to explore the answers to our research questions using applied research methodologies.

Since we are using an existing dataset, the data can be termed as a secondary source. This implies that we have no control over how the data was generated. Hence, we would use data pre-processing to ensure that the data suits the analysis. Next, our results will be based on experiments conducted on the dataset. Working with quantitative methodology will ensure that our results will be reproducible and standardised. 

\subsection{Note on Dataset}
The UCI Adult dataset was compiled by Barry Becker, who extracted the data from the US 1994 Census database \cite{Lundberg}. This dataset was given to Ron Kohavi \cite{Kohavi1994}, who used it to study the effectiveness of a new machine learning algorithm he proposed called the NBTree. After the completion of Kohavi's paper in 1996, the dataset was donated and is now hosted by the Machine Learning Group at UC Irvine. It is also called the 'Income' or 'Census' Dataset. Since then, the dataset has been referenced in roughly 50 academic papers. In many of these papers, researchers studied the performance of augmenting existing classification machine learning algorithms, such as SVMs and K-NNs, by boosting, partitioning, and squashing. Due to the dataset's numerous citations, it is well-known in the Data Mining and Machine Learning community. The dataset contains about 34,000 samples. Each sample has a set of demographic attributes of people, such as their age, education, and marital status. It categorises their income as above or below \$50,000. 

Other researchers support the view of choosing the Adult dataset \cite{Mehrabi2021} . The authors in Explainable and Non-explainable Discrimination in Classification" \cite{Kamiran2013}  state that it is a widely used dataset in the fairness domain. Our work can be extended to analogous datasets easily. These datasets include US Census Demographics Data \cite{USCensus}, India Census Data \cite{Census}, and Covid-19 Surveillance data \cite{Covid} . 

\subsection{Procedure}
It is proven that data is biased by nature due to the cognitive bias of human brains \cite{Taniguchi2018}. Therefore, evaluating the potential bias of machine learning models would be valuable for better model explainability. Thus, we will detect biased attributes with a certain confidence level and determine which attribute is causing the bias. 

Since the target attribute 'wage' uses continuous values, we use regression algorithms to predict it. The algorithms used are a combination of linear regression, polynomial regression and other open-sourced libraries such as LightGBM \cite{lightgbm}, XGBoost \cite{xgboost_2016}, Catboost \cite{catboostdorogush2018}, and Tabnet \cite{TabNet}. Note that any other machine learning algorithm can be used for the experiments. 

Stacking models involves combining the predictions from multiple machine learning models on the same dataset. There are two practical approaches to stacking - bagging and boosting. Bagging allows averaging multiple similar models with a high bias to decrease the variance. Boosting builds multiple incremental models to reduce the bias while keeping variance small. Stacking can be seen as exploring a space of different models for the same problem. The idea is that one can attack a learning problem with varying models capable of learning some part of the problem but not the whole space. Hence one can build multiple learners and use them to make an intermediate prediction, i.e. one prediction for each learned model. Then, one adds a new model that learns the same target from the intermediate predictions. This final model is said to be stacked on top of the others. Thus, one may improve the overall performance and often end up with a better model than any individual intermediate model. At this stage, we define notations and concepts such as Alternation Functions, Potentially Biased Attributes and KL Divergence as defined in the previous literature \cite{Taniguchi2018}. 

\subsection{Notations and definitions}

\subsubsection{Notations}
The dataset is represented by $D$. We assume that it has $n$ samples and $m$ attributes. $D$ can be shown as a set of attribute vectors $ \{ a_{1},\ a_{2},\ \ldots,\ a_{j}\ ,\ \ldots, a_{m} \} $. The actual target values are denoted by the vector $ \{y = \{ y_{1},\ y_{2},\ \ldots,\ y_{j}\ \ ,\ \ldots,\ y_m \}$. 
The predicted target is shown by $\hat{y}$. The equation $f(D) \rightarrow\ \hat{y} $ represents a model $f$ which takes the dataset as input and assigns each data sample to a specific target $\hat{y_{i}}$ .

\subsubsection{Alternation Functions}
Alternation is a function that alternates between attributes' values. It swaps the values every time the function is applied. In other words, alternation takes an attribute and changes the instance's identity. We aim to check the dependency of the model on the selected PBA. We denote the alternation function as $\varphi(\cdot)$. The function takes a dataset as an input and returns the alternative dataset. The alternative dataset is the dataset with the alternative values of a specific attribute. 
\begin{equation*}
    \varphi(D)= \{a_1,\ a_2,\ \ldots,\  -a_j  , \ldots ,a_m \}
\end{equation*}

It can be seen that $\varphi(\cdot)$ is a function that switches the protected attribute's values such that ${a_j}$ becomes $-{a_j}$. Assuming that the protected attribute (${a_j}$)  is gender, the vector ($ -{a_j}$) represents where the female values become male, and vice versa.

\subsubsection{Potentially Biased Attributes}
Without loss of generality, we assume that $a_j$ is a categorical attribute. We can say that $a_j\in D$ is a potentially biased attribute if $f(D)\neq f(\varphi(D))$. \\
Here, $\varphi(D)$ is an alternation function and $\hat{y_{\varphi}}$ represents the alternative predictions.

For example, let $D$ be a dataset of n applicants, and ${ a_j}$ is a gender attribute, while $y$ represents whether the applicant is qualified or not qualified. If the prediction changes by only changing the gender and maintaining the values of the remaining attributes, then the attribute $ {a_j}$ is said to be a Potentially Biased Attribute (PBA). It can be said that a model is considered biased if it is dependent on one or more PBA given the class label \cite{Jiang2020}

\subsubsection{KL Divergence}
Kullback-Leibler (KL) divergence is used to estimate the distribution difference between the actual target $(y)$ and predicted target $\hat{y}$ for each value in PBA \cite{Legrand2018} . Assuming we have the densities $p$ and $q$, we calculate the divergence between them using the following equation
\begin{equation*}
     D_{KL}(p||q ) = \int p(x) \log{\dfrac{p(x)}{q(x)}} \ dx
\end{equation*}

Considering $p$ and $q$ are derived from normal distributions with means $ \mu_1, \mu_2$ and standard deviations $\sigma_1, \sigma_2$, the KL divergence can be represented as the following: 
\begin{equation*}
    D_{KL}(p||q ) = \log{\dfrac{\sigma_2}{\sigma_1}} + \dfrac{\sigma_{1}^{2} + (\mu_1 - \mu_2)^{2}}{2 \sigma_{2}^{2}} - \dfrac{1}{2} 
\end{equation*}

If the result of $D_{KL}$ is zero, it can be said that the two distributions are identical. This implies that this attribute causes no bias. Otherwise, if the distributions differ, bias may be reported. The larger the result $D_{KL}$ is, the difference between the distributions is more significant, indicating greater bias.

While we use the regression algorithms on the dataset, we must ensure consistency in results. Several Cross-Validation(CV) techniques such as Holdout, K-fold, Stratified k-fold, Monte Carlo and Leave-one-out method have been developed to get consistent results. The documentation of the popular ML library scikit-learn \cite{Pedregosa} mentions that the k-fold cross-validation scheme can be used to evaluate the estimator's performance. One advantage of the cross-validation analysis is that it avoids the problem of overfitting. It ensures that we get generalisable results on a test set. It can be imported from the model selection in module sci-kit learn. We apply the k-fold cross-validation model selection technique to obtain consistent results. CV is known for its statistical ability to generate less biased datasets, known as data folds or folds. 

For each fold, we train the model using 90\% of the data, while the remaining 10\% is used for testing. We predict the target using the model in each fold. After that, we apply the alternation function and predict the target variable, i.e. wage, again for the same folds. Unlike training folds, where each sample will appear in training folds $k-1$ times, each sample will be used for testing in just one fold. After predicting the wage in each fold, we predict it again using the same instance but by alternating the PBA. Once this process is completed, we repeat the experiment with a stacked model and check for variations in results.

Our evaluation intends to find the dependency of the prediction on the protected attribute; and its effect when using a stacked model. Suppose the model predicts significantly different densities after the alternation function is applied. In that case, this is an indicator that the prediction depends on that PBA. We consider the density of the predicted class label of the original dataset $D$ to be the original distribution. It is denoted as $p$. In comparison, the density prediction of the alternated dataset is denoted as $q$. $D_{KL}$ quantifies how much $p$ deviates from $q$. In other words, $D_{KL}$ evaluates the divergence between the density of predicted class label $p_{\hat{y}}$ and the density of the predicted class label $q_{\hat{y}}$.

Overall, the methodology can be summarised as follows:
\begin{enumerate}
    \item Train the model $f(\cdot)$ on the dataset $D$
    \item Predict the class label for each data point in $f(D)\rightarrow\  \hat{y}$. The target distributions for males and females are represented by $p_{\hat{y}_m}$ and 
    $p_{\hat{y}_f}$ respectively
    \item Apply the alternation function on the gender attribute $\varphi(D)\rightarrow D_{\varphi}$
    \item Train the model $f(\cdot)$ on the alternative dataset $D_{\varphi}$
    \item Predict the alternative predicted label $f(D_{\varphi}) \rightarrow \hat{y}_{\varphi}$. The alternated target distributions for males and females are represented by $q_{\hat{y}_{\varphi f}}$ and $q_{\hat{y}_{\varphi m}}$ respectively. 
	\item Evaluate KL Divergence between the distributions of $\hat{y}$ and $\hat{y}_{\varphi}$ for each gender distinctively across all folds. That is, evaluate $ D_{KL}(p_{\hat{y}_m} || q_{\hat{y}_{\varphi f}} ) $ and $ D_{KL}(p_{\hat{y}_f} || q_{\hat{y}_{\varphi m}} ) $
	\item The difference between the $D_{KL}$ with respect to gender represents the bias. The larger $D_{KL}$is, the larger the bias will be.
	\item Repeat steps 1-7 using a stacked ML model instead of a single one. Ensure that each regression model is given appropriate weight while implementing bagging.
\end{enumerate}

\begin{figure*}[]
    \begin{center}
    \includegraphics[width=6in]{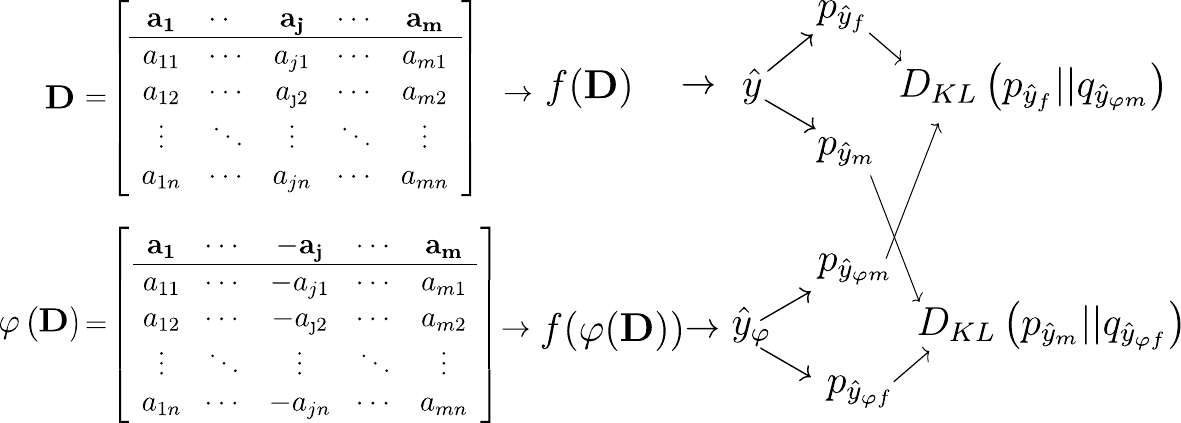}
    \end{center}
    \caption{\small A summary of the experimental procedure as given by \citet{Alelyani2021}. We begin with an initial dataset, on which the algorithm is applied to get predictions. The predictions are split according to attributes such as predictions for males ($m$) and females($f$). A similar process is then applied to the alternated dataset. Finally, we compare the KL Divergence scores to check for the dependency of the result on a PBA}
    \label{fig: full_process}
\end{figure*}

The process can be shown pictorially as in Fig. \ref{fig: full_process}

\subsection{Feature Study and Selection}
We chose selected categorical variables as the features from the dataset. These include gender, migration status, race, country of birth, citizenship status, marital status, and the class of work. Since the original Income dataset is used to classify the annual income as more than \$50,000 or not, we drop this attribute in our experiments because it may contribute to data leakage. Data leakage refers to training an ML model on data which has information about the target variable. It results in an unreliable model once deployed. To avoid data leakage, we drop the following attributes from the dataset 
\begin{itemize}
    \item industry code
    \item reason for unemployment
    \item employment status
    \item capital gains
    \item capital losses
    \item dividends from stocks
    \item tax filler status
\end{itemize}
Dropping such features may result in a less accurate wage prediction model. However, our objective is only to observe the change in results once we apply alternations. Hence, we only reject features which may directly lead to data leakage.

Another essential step in the feature selection process is the removal of outliers values. We utilise Box and Whisker plots for visualising the attributes. A key advantage of using these plots is their easy readability. Based on the distribution of values, we drop samples which have values beyond the \nth{99}  percentile of the entire range. The sensitivity of a model on outliers has been demonstrated by Acuna \cite{Acuna2004}. Outlier detection and removal is one of the most commonly used techniques for data cleaning. 

 \begin{figure*}[]
  \begin{center}
    \includegraphics[width=4.5in]{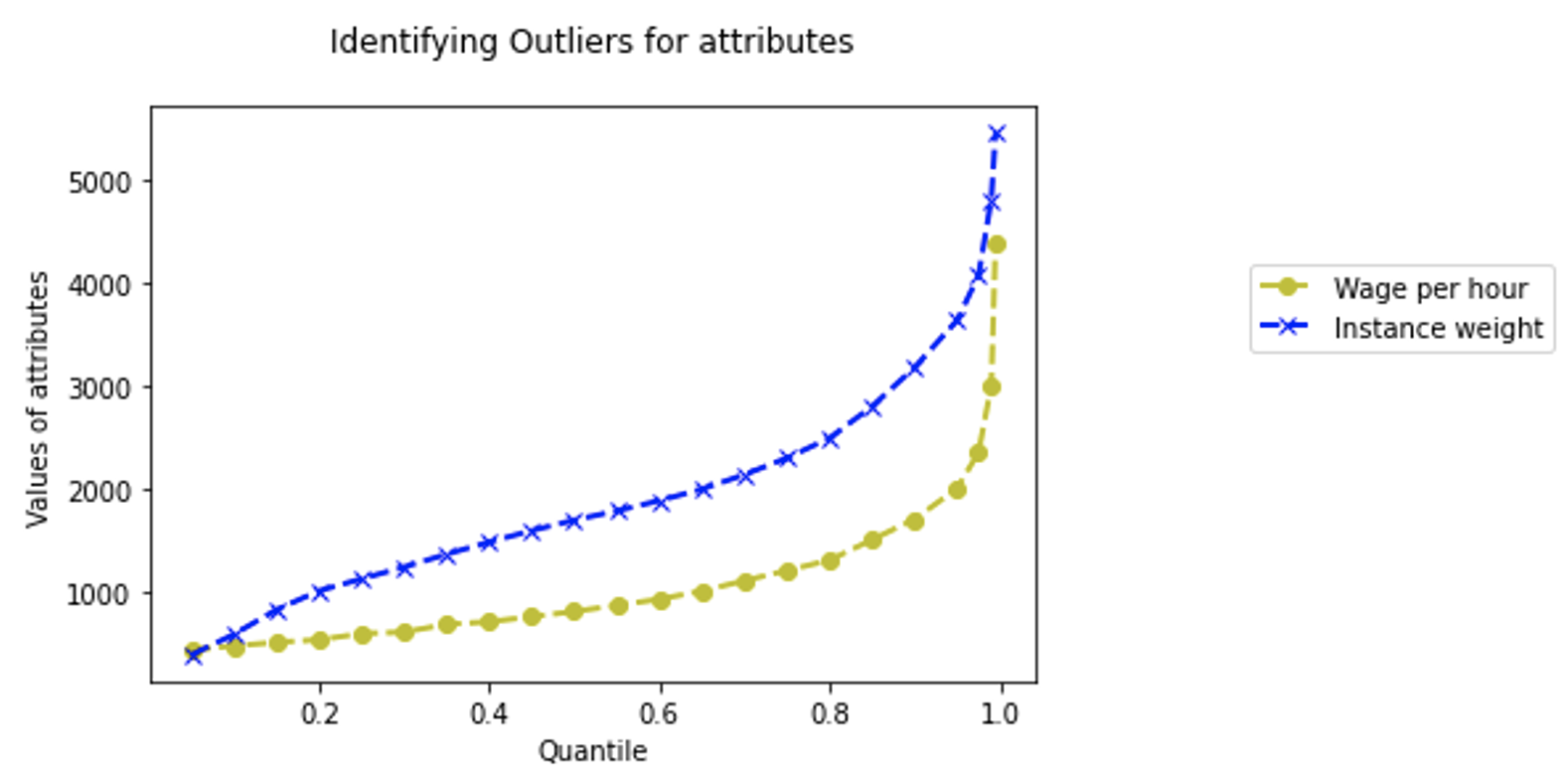}
  \end{center}

  \caption{\small Outlier detection by plotting the distributions by quantile value}
  \label{fig: outlier_detection}
\end{figure*}


\begin{table*}[!htbp]
    \centering
    \begin{tabular}{cccccc}
        \toprule
        \multirow{2}{*}{\textbf{Measure}} & \multicolumn{5}{c}{\textbf{Attributes}} \\
         & age & \begin{tabular}[c]{@{}c@{}}detailed industry\\ recode\end{tabular} & \begin{tabular}[c]{@{}c@{}}detailed occupation\\ recode\end{tabular} & \begin{tabular}[c]{@{}c@{}}wage per\\ hour\end{tabular} & \begin{tabular}[c]{@{}c@{}}instance\\ weight\end{tabular} \\
         \midrule
        count & 10187 & 10187 & 10187 & 10187 & 10187 \\
        mean & 36.22 & 30.848 & 26.804 & 838.23 & 1760.7 \\
        std & 13.28 & 12.423 & 10.333 & 347.8 & 926.93 \\
        min & 15 & 1 & 1 & 20 & 62.37 \\
        25\% & 25 & 27 & 19 & 550 & 1107.3 \\
        50\% & 35 & 33 & 29 & 760 & 1675.4 \\
        75\% & 45 & 41 & 35 & 1050 & 2274.3 \\
        max & 90 & 50 & 45 & 1745 & 4676.5 \\
        \bottomrule
    \end{tabular}
    \caption{\small Statistical information describing numeric attributes}
    \label{tab:info table}
\end{table*}

\subsection{Preprocessing}
Before we begin processing the dataset, cleaning the data with specific pre-processing techniques is essential \cite{Chakrabarty2018}. This includes the following essential steps:

\subsubsection{Handling missing values}
The dataset contains missing values for categorical features such as occupation and country of birth. The missing values have been dealt with by dropping the corresponding sample from the data. The missing values are present either as NaN values or marked as 'Not in Universe'. Since the relative number of samples having such missing values is negligible compared to the number total values, dropping those samples will ensure we are not leaving out important information. According to \citet{Rahman2013} , extra information about the data can be extracted from the missing values. However, our dataset's low proportion of missing values allows us to drop the corresponding samples.

\subsubsection{Encoding of Categorical features}
We use the popular Label Encoding technique for dealing with categorical features. It can be accessed and implemented using the scikit-learn library under the pre-processing module. Here, each label is assigned a unique integer based on the alphabetical ordering of the variable. For instance, in the gender attribute, all males are given the value $0$, while females are represented by the value $1$. It should be noted that all the categorical variables we are dealing with have no more than two possible categories. For example, the variable concerning Citizenship Status has only two categories - (i) US citizen and (ii) Not US Citizen. This results in the columns having binary values after label encoding. Since we would apply alternations to these binary variables, the alternated data would also have a binary form for each attribute \cite{Pedregosa}. Since the values of variables are binary, one-hot encoded columns will only add redundant columns to our data, which is undesirable. Additionally, to avoid the curse of dimensionality \cite{Chen2009} , using the One-Hot Encoding technique is conveniently avoided.

\subsubsection{Shuffling and Splitting}
The dataset is randomly sampled and shuffled. It guarantees that all the categories of different attributes are included in both the training and validation sets. As mentioned earlier, the k-fold cross validation is used for splitting. Each iteration uses 90\% of the data for training and the rest 10\% for testing.

\subsection{Learning Algorithms}
We follow the step-by-step procedure mentioned earlier while describing the Experimental Procedure. We use the following ML regression algorithms -
\begin{enumerate}
    \item XGBoost (XGB)
    \item LightGBM (LGBM)
    \item Gradient Boosting Trees (GB)
    \item Random Forests (RF)
    \item Linear regression (LinearR)
    \item Lasso regression (LassoR)
\end{enumerate}
 To choose the best set of hyperparameters for the models, we use the open-source Optuna Library \cite{Akiba2019}. The summary of Optuna search of the Gradient Boosting Regressor is below

 \begin{figure*}[]
  \begin{center}
    \includegraphics[width=5.5in]{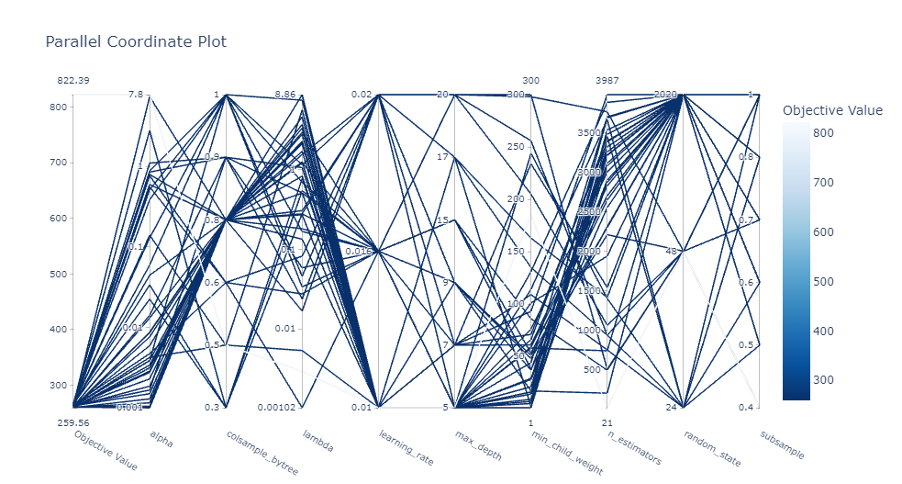}
  \end{center}

  \caption{\small Summary of Hyperparameter Tuning when using Optuna. Note that the algorithm is adjusted to minimise the 'Objective Value' or the loss function. Optuna narrows down searches using hyperparameters which give a better value of the loss function. The hyperparameters tuned are (i) alpha (ii) number of columns per sample by each tree, (iii) lambda (iv) learning rate (v) maximum depth (vi) minimum weight of child (vii) number of estimators (viii) random state and (ix) subsample}
  \label{fig: optuna}
\end{figure*}

\subsection{Implementation}
The Data  Pre-processing and  Model  Development steps are done using Python's open-sourced Scikit-Learn Library \cite{scikit-learn}. We run the ML models on a Google Colaboratory \cite{Bisong2019} CPU instance Intel(R) Xeon(R) single-core CPU @ 2.20GHz, with 8 GB RAM. The Visualisations are made using Python's Plotting Libraries, Matplotlib \cite{Hunter2007} and Seaborn \cite{Waskom2021}. Optuna is used to tune the hyperparameters.


\section{Experimental Findings}
We aim to explain the effect of every attribute on the predicted wage. It is important to stress that the predictions are made for each sample in the test set obtained during the k-fold cross validation process. The experiment was conducted with 15-folds cross-validation. We show the results for two attributes - gender and country of citizenship using the XGB Regressor model. The complete results for the two attributes are given in the following figures and tables

 \begin{figure*}[]
  \begin{center}
    \includegraphics[width=5.0in]{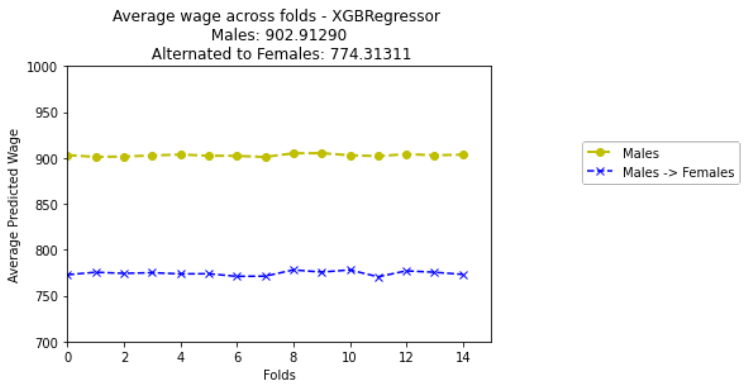}
    \includegraphics[width=5.0in]{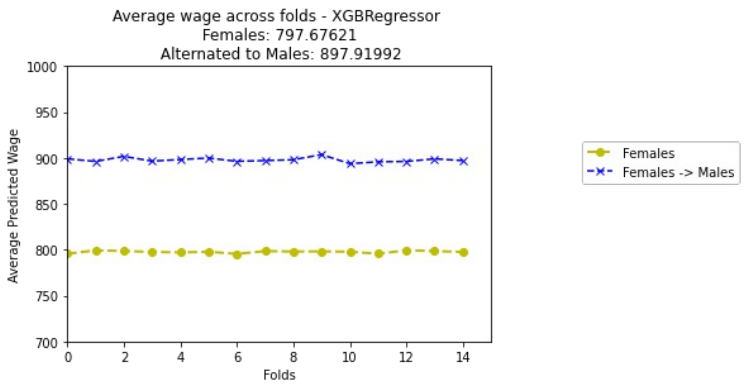}
    \includegraphics[width=5.0in]{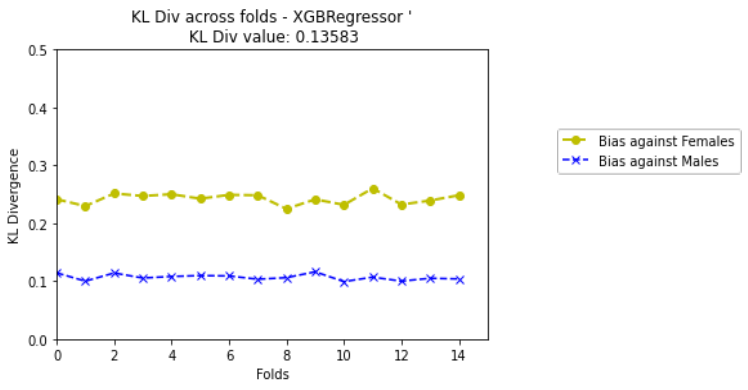}
  \end{center}
  \caption{\small Plots (a) and (b) represent the average predicted wage across folds and (c) shows KL Divergence Across folds, when the XGB Regressor is used}
  \label{fig: xgb_males}
\end{figure*}

According to Alelyani, plotting the KL divergence across folds measures the bias. Figure 5 .a shows that the average predicted wage for males is about \$900. When they are alternated to females, this prediction drops to about \$770. Similarly, Figure 5 .b shows how the predicted wage increases for females when they are alternated to males. Figure 5 .c quantifies the results. It clearly shows a consistently higher bias against females than males. Note that the graphs show results from the XGB Regressor, which is similar to the results of other models we used.
A full table with the results for the gender attribute is given in Table \ref{tab:gender table} (all values are in US Dollars \$) 

\begin{table*}[!htbp]
    \centering
    \begin{tabular}{ccccc}
        \toprule
        \textbf{ML Model} & \textbf{Mean values} & \textbf{Predictions} & \textbf{Alternated Predictions} & \textbf{Average KL Divergence} \\
        \midrule
        \multirow{2}{*}{XGB} & Males & 902.9128 & 774.3131 & \multirow{2}{*}{0.13582} \\
         & Females & 797.6762 & 897.9199 &  \\
         \hline
        \multirow{2}{*}{LGBM} & Males & 904.2705 & 775.9298 & \multirow{2}{*}{0.14007} \\
         & Females & 796.4586 & 898.7639 &  \\
         \hline
        \multirow{2}{*}{GB} & Males & 904.7893 & 775.6511 & \multirow{2}{*}{0.14418} \\
         & Females & 798.3236 & 903.7192 &  \\
         \hline
        \multirow{2}{*}{RF} & Males & 908.3169 & 770.6641 & \multirow{2}{*}{0.19390} \\
         & Females & 803.1593 & 901.5988 &  \\
         \hline
        \multirow{2}{*}{LinearR} & Males & 902.6762 & 758.1122 & \multirow{2}{*}{0.05956} \\
         & Females & 793.3388 & 937.9028 &  \\
         \hline
        \multirow{2}{*}{LassoR} & Males & 901.6486 & 758.8464 & \multirow{2}{*}{0.06137} \\
        & Females & 793.7822 & 936.5844 &  \\
        \bottomrule
    \end{tabular}
    \caption{\small Results for Gender Attribute. The actual wage distribution for Males is \$897.6978; for Females it is \$805.9626}
    \label{tab:gender table}
\end{table*}

The actual distribution of average income in the test set for Males is \$897.69. It can be seen that each of the six ML models predicts wages close to this value. For example, XGB  predicts \$902.91, and GB predicts \$904.78. It indicates that the models have been trained well since they closely predict the average wage. A similar argument can be made for the Females column. The actual value from the distribution is \$805.95, while the LGBM predicts \$796.45, and the RF predicts \$803.15.

The most important observations are from the Alternated predictions of our ML models. For instance, consider the XGB Regressor: The average prediction for Males is \$902.91, which falls sharply to \$774.31 after alternating the Male attribute to Female. Moreover, this alternated prediction is close to the average wage predictions for Females. It shows how severely the gender impacts the average predictions for Males and Females. It indicates that the trained model has learnt that Males should make significantly more money than Females, considering all other attributes are identical. This observation is reinforced when we consider the Females column. Females' actual distribution of wages is \$805.95, and their predicted value is \$797.67. However, once the alternation is applied, i.e., females are switched to males, the predicted wage rises sharply to \$897.91. This result lies close to the predicted wages for Males. The gender attribute influences the prediction of the wage. There is a clear indication of bias against females. 

To quantitively measure how different the predictions and the alternate predictions are, we calculate the average KL divergence across all the 15 folds. The values of KL divergence for XGB, LGBM, GB, and RF regressors are more than 0.13, while the same value for Linear and Lasso Regressors is around 0.05. These values indicate that the tree-based models have more strongly learnt that Males should have a higher wage than their female counterparts. Since the Linear Regressor and Lasso Regressor are mathematically simpler ML models \cite{kutschkem2018} , we argue they are not complex enough to gain deeper insights into the dataset. Hence, these two models may not be given as much weightage as the others when we proceed to stack many ML models into one. 

It can be noted that since all XGB, LGBM, GB, and RF regressors are tree-based models, they are relatively similar in their work. Hence, one may expect similar results from all the models once training is done. 

\begin{figure*}[!htbp]
  \begin{center}
    \includegraphics[width=5.0in]{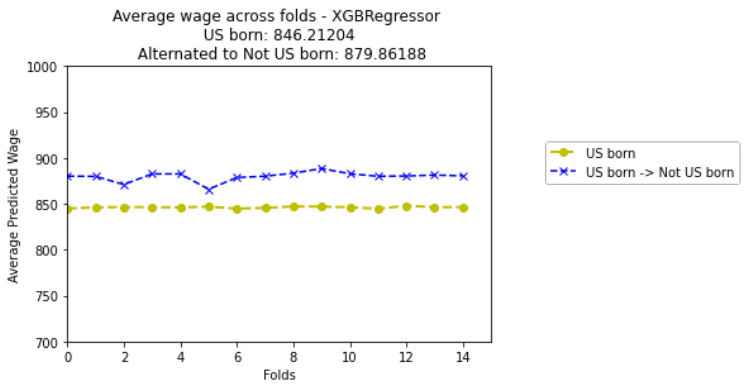}
    \includegraphics[width=5.0in]{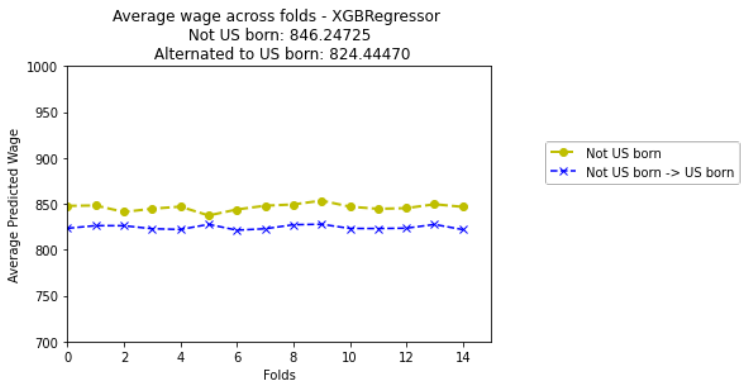}
    \includegraphics[width=5.0in]{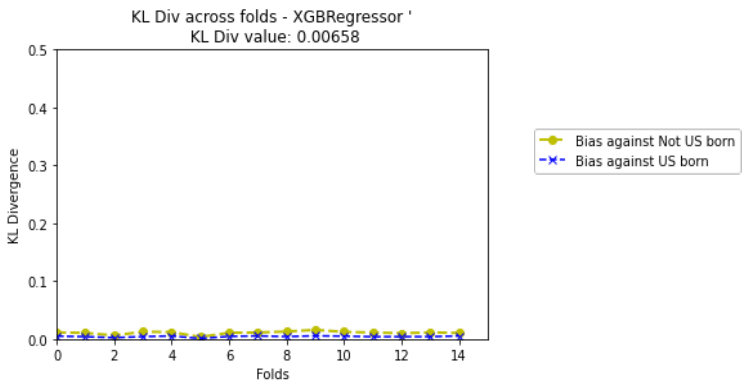}
  \end{center}
  \caption{\small Plots (a) and (b) represent the average predicted wage across folds for the Country of Birth Attribute,  and (c) shows KL Divergence Across folds, when the XGB Regressor is used. }
  \label{fig: xgb_us}
  
\end{figure*}

Following are the results plotted for another attribute - Country of Birth, shown in Fig. \ref{fig: xgb_us}

From Figure \ref{fig: xgb_us}.a and Figure \ref{fig: xgb_us}.b, our model may seem to prefer people who are not born in the US for a higher wage. When US born people are alternated to non US born, their predicted wage increases, whereas when the alternation is carried the other way, the predicted wage decreases. it must be underlined that even though the plots in figures a and b are distinct, we cannot measure this difference by visual inspection. Figure \ref{fig: xgb_us}.c shows how KL divergence varies across folds. A higher value of KL divergence represents a greater tendency of the algorithm to favour a specific value of the variable, i.e. more bias. Clearly, the amount of bias against either of the two groups is very close to zero, representing negligible bias. Hence the ML models do not prefer any one group in particular.

A full table with the results for the Country of Birth attribute is given in Table \ref{tab:country table} (all values are in US Dollars \$) -  

\begin{table*}[!htbp]
    \centering
    \begin{tabular}{ccccc}
        \toprule
        \textbf{ML Model} & \textbf{Mean values} & \textbf{Predictions} & \textbf{Alternated Predictions} & \textbf{Average KL Divergence} \\
        \midrule
        \multirow{2}{*}{XGB} & US Born & 846.2120 & 879.8619 & \multirow{2}{*}{0.00657} \\
         & Not US Born & 846.2472 & 824.4447 &  \\
         \hline
        \multirow{2}{*}{LGBM} & US Born & 846.2548 & 883.2602 & \multirow{2}{*}{0.00755} \\
         & Not US Born & 845.4938 & 820.9484 &  \\
         \hline
        \multirow{2}{*}{GB} & US Born & 846.9861 & 874.2755 & \multirow{2}{*}{0.00420} \\
         & Not US Born & 851.8441 & 833.8565 &  \\
         \hline
        \multirow{2}{*}{RF} & US Born & 851.3212 & 866.5876 & \multirow{2}{*}{0.00269} \\
         & Not US Born & 855.0529 & 851.1423 &  \\
         \hline
        \multirow{2}{*}{Lin. Reg} & US Born & 842.5040 & 871.2517 & \multirow{2}{*}{0.00167} \\
         & Not US Born & 856.3671 & 827.6194 &  \\
         \hline
        \multirow{2}{*}{Lasso Reg} & US Born & 842.9185 & 844.7389 & \multirow{2}{*}{0.00001} \\
         & Not US Born & 849.6640 & 847.8436 & \\
        \bottomrule
    \end{tabular}
    \caption{\small Results for Country of Birth Attribute. The actual wage distribution for US Born citizens is \$850.8963; for Non-US born is \$822.1075}
    \label{tab:country table}

\end{table*}

Table \ref{tab:country table}  shows results when the same process is applied to the Country of Birth attribute. We can observe a general trend of the alternated predictions for US Born citizens being higher than actual predictions. Analogously, the Alternated predictions for Non-US Born citizens are lower than the actual predictions. At first glance, it may seem that there exists a bias against US Born citizens. Because the average predictions for wage increases simply by changing their country of birth to a Non-US country. However, the very small KL divergence values suggest that there is actually significantly less bias for or against any US Born/Non-US Born citizens. This example demonstrates the importance of calculating a measurable value for checking bias. Suppose, instead of calculating the KL Divergence value, we directly try to eyeball the results from figures b and c. In that case, it can easily lead to an incorrect conclusion that a bias exists for this attribute.

We present is a concise Table \ref{tab:all PBA} with Average KL Divergence Values for every PBA tested
\begin{table*}[!htbp]
    \centering
    \begin{tabular}{cccccccc}
    \toprule
    \multirow{2}{*}{\textbf{Algorithm}} & \multicolumn{7}{c}{\textbf{Attribute(s)}} \\
     & Gender & Migrants & \begin{tabular}[c]{@{}c@{}}Birth \\ country\end{tabular} & Citizenship & \begin{tabular}[c]{@{}c@{}}Living \\ Together\end{tabular} & \begin{tabular}[c]{@{}c@{}}Class of \\ worker\end{tabular} & Race \\
     \midrule
    XGB & 0.13582 & 0.00066 & 0.00657 & 0.01107 & 0.00276 & 0.00007 & 0.00076 \\
    \hline
    LGBM & 0.14007 & 0.00065 & 0.00755 & 0.01179 & 0.00141 & 0.00338 & 0.00260 \\
    \hline
    GB & 0.14418 & 0.00031 & 0.00420 & 0.00325 & 0.00333 & 0.00989 & 0.00101 \\
    \hline
    RF & 0.19390 & 0.00015 & 0.00269 & 0.00193 & 0.00160 & 0.00181 & 0.00088 \\
    \hline
    LinearR & 0.05956 & 0.00012 & 0.00167 & 0.02994 & 0.15450 & 0.06551 & 0.00222 \\
    \hline
    LassoR & 0.06137 & 0.00009 & 0.00001 & 0.00930 & 0.15404 & 0.06230 & 0.00143 \\
    \bottomrule
    \end{tabular}
    \caption{\small Average KL Divergence values for all PBAs}
    \label{tab:all PBA}

\end{table*}

We observe how the Gender column in Table \ref{tab:all PBA} shows a significant value of KL divergence. Other columns have values very close to zero. The Linear and Lasso regressors show deviant behaviours for the Living Together Attribute. However, since the other regressors do not show consistent results, it cannot be inferred that this column can be considered as a PBA. Statistically, we may apply Hypothesis tests \cite{Banerjee2009} to check if a value of KL Divergence is negligible or not. However, we exclude the use of the Hypothesis tests and other statistical approaches because they are beyond the objective of this work.

\begin{figure*}[]
  \begin{center}
    \includegraphics[width=5.0in]{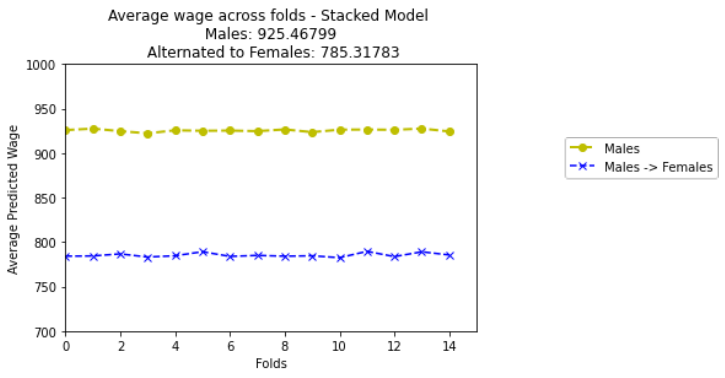}
    \includegraphics[width=5.0in]{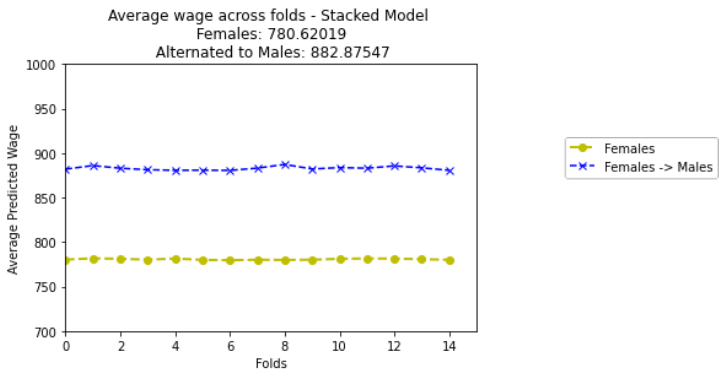}
    \includegraphics[width=5.0in]{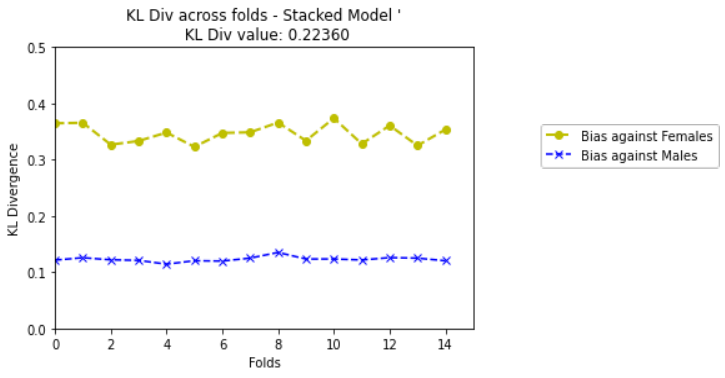}
  \end{center}
  \caption{\small The results when we use a stacked model on the Gender Attribute. As earlier, we plot the average wage across folds for both Males and Females; and the KL Divergence values for the distributions}
  \label{fig: stacked}
  
\end{figure*}

Considering that only the Gender attribute shows a significant value of KL Divergence, we run the stacked model only on the Gender Column to check if it makes a difference. As mentioned in the previous observations, the weightage given to the Linear and Lasso Regression Models is low. In contrast, the weights for other models are higher. Our stacked model implements the Bagging approach, wherein we take a weighted average of the results from individual models. For every four units of weight given to the tree-based models, we give one unit of weight to the Linear and Lasso Regression models. The results after implementing the stacked model on the Gender attribute are given in Fig. \ref{fig: stacked} -

We observe how the predictions are more biased against females than against males. The stacked model performs similarly on the Gender Attribute compared to individual models tested earlier. This can be attributed to the fact that all models were tree-based, except Linear and Lasso Regressors. Hence they are broadly similar in their work and results. The lower weightage we give to the Linear and Lasso Regressors in the stacked model also plays a role in the stacked model performing similar to the tree-based models.



\begin{table*}[!htbp]
    \centering
    \begin{tabular}{ccccc}
        \toprule
        \textbf{Algorithm} & \textbf{Mean values} & \textbf{Predictions} & \textbf{Alternated Predictions} & \textbf{Average KL Divergence} \\
        \midrule
        \multirow{2}{*}{Stacked Model} & Males & 925.4679 & 785.3178 & \multirow{2}{*}{0.2236} \\
         & Females & 780.6201 & 882.8754 &  \\
        \bottomrule
    \end{tabular}
    \caption{\small Results when using a Stacked model on the Gender Attribute. The actual Distribution for Male is \$955.1808; for Females is \$782.6446}
    \label{tab:stacked table}
\end{table*}


\section{Conclusion}
This report proposed the application of  Ensemble  Learning to the UCI Adult Dataset. We have predicted the measurable values for bias against multiple categorical variables, using alternate functions defined in literature \cite{Alelyani2021}. We used multiple ML models for the task and quantified the results. Finally, the gender attribute, which has the most significant amount of bias, was inspected using a Stacked ML model. 
One may become successful in achieving an overall better set of results by using hybrid models. This can be achieved by including Machine Learning and Deep Learning or applying many other advanced pre-processing techniques without further depletion in the results. As mentioned by \citet{Chakrabarty2018},  we can reject the pessimism of \citet{Corbett2017}, who had been worried that the cost of fairness reduced learner performance. 

We have quantified bias levels against a group based on their gender, race, citizenship etc. However, it should be noted that this study may be extended further in multiple directions. For instance, instead of KL divergence as suggested by Alelyani3, we may choose Jensen-Shannon Divergence, which can be viewed as the symmetrised version of the KL divergence \cite{Wang2018}. Another substitute may be Jeffrey's divergence, which has been used in various applications \cite{Legrand2018} . As mentioned in the report, there is a scope for applying statistical tests like Hypothesis tests and using confidence intervals and p-values. These may be applied for further checking which values of bias are significantly large to be studied further and which are not. This step may impact the choice of ML models used while using the stacked model. One may include varying types of regression models in the stacked models, leading to a better quality of results, as seen in earlier studies \cite{Hansen2020, Lim2022} . These different models may include Deep Learning models, which use Neural Networks in their work. The Keras library in Tensorflow provides easy-to-implement frameworks for deploying such DL models for tabular data \cite{Chollet2015} . Similarly, libraries such as TabNet \cite{TabNet}  based on the Pytorch framework can be used. In a general sense, there is scope to engineer domain-specific features from the given data instead of blindly using optimisation methods and ML algorithms. 

Increasingly, software is making autonomous decisions in case of criminal sentencing, approving credit cards, and hiring employees. These decisions show bias and adversely affect certain social groups like those defined by sex, race, age, and marital status. In such cases, it becomes critical that the software making the autonomous decisions does not prefer a particular group in an unfairly biased manner. Even though such ML models are not necessarily programmed to give biased results, it becomes increasingly crucial that the developers test the models thoroughly for any such biases, as mentioned in the report. Rigorous testing with real-world data will ensure that the model behaves 'fairly'. Care must also be taken to ensure that models that learn from their outputs (i.e. reinforcement learning-based) do not learn parameters that increasingly lead to higher bias against a particular group.


\section{Future Work}

This study serves as a stepping stone to a multifaceted realm of possibilities for advancing bias mitigation in machine learning. To enhance the quality of results and deepen our understanding, the adoption of advanced pre-processing techniques and hybrid models stands as a promising avenue. These approaches involve the fusion of machine learning and deep learning, aiming to circumvent potential performance reduction, an apprehension previously voiced within the research community. Integrating neural network-based models, such as those facilitated by the Keras library within the TensorFlow framework, provides a rich environment for predictive models in the context of tabular data. Additionally, exploring PyTorch-based solutions like TabNet holds potential, opening doors to novel architectures in bias detection and elimination.

While Kullback-Leibler divergence suffices as an effective measure for quantifying bias, an extension of this research may delve into the application of alternative divergence metrics. Notably, Jensen-Shannon Divergence, a symmetrized variation of KL divergence, and Jeffrey's Divergence, which boasts multiple applications, may provide deeper insights. Statistical hypothesis tests and the incorporation of confidence intervals and p-values represent further avenues for investigating the significance of bias levels, potentially influencing the selection of machine learning models within ensemble learning. Subsequently, one can extend the ensemble models to encompass a diverse assortment of regression models, going beyond the constraints of this study, as previous literature has demonstrated the positive impact of model diversification. 

The ongoing developments in data-driven autonomy across domains necessitate vigilant scrutiny for bias. Further research must rigorously test these autonomous systems to ensure fairness and impartiality across diverse sociodemographic groups. Exploring emerging methodologies and diversified feature engineering from data provide a broader spectrum of opportunities, ultimately fostering fairness and inclusivity in the rapidly evolving machine learning landscape.



\section{The Acknowledgements}
This work is the result of a year of project-specific work spent at IIT Madras for the Young Research Fellowship (YRF) project. I had the opportunity to work closely with Professor Satya Sundar Sethy, whom I thank sincerely for giving me my chance, and above all, for having guided, supported, and encouraged me throughout the project. His constant insights into research writing have ensured that the project remains on track. The fantastic team of Professors for the YRF Program, which includes Prof. Sankaran Aniruddhan, Prof. Sachin Gunthe, and Prof. Preeti Aghalayam and Sumani ma'am, who made the entire program not only a close-knit team but a united family. My classmates and friends, with whom I had insightful discussions on topics of interest. Huge credit also goes to my Career and Life Coaching (CLIC) mentor - Mr Krishnamurti Rao, for always being my sounding board in every situation and providing answers from an excellent perspective. Dr Salem Alelyani, who has been very supportive and ready to give his time and share his knowledge on this topic. \\


\bibliographystyle{unsrtnat}

\bibliography{refs}  

\end{document}